\documentclass{amsart}
\usepackage{amssymb,latexsym}
\usepackage{graphicx}

\newcommand{\de}{\mathrm{d}}
\newcommand{\e}{\mathrm{e}}
\newcommand{\nn}{\nonumber}
\newcommand{\ug}{\!\!\!\!&=&\!\!\!\!}

\begin{document}

\title{On Ramanujan Master Theorem}
\author{D. Babusci, G. Dattoli}
\address{INFN - Laboratori Nazionali di Frascati, via E. Fermi 40, I-00044 Frascati.}
\address{ENEA - Centro Ricerche Frascati, via E. Fermi 45, I-00044 Frascati.}

\begin{abstract}
In this short note we use the umbral formalism to derive the Ramanujan Master Theorem and discuss its
extension to more general cases.
\end{abstract}

\maketitle

The Ramanujan Master Theorem (RMT) \cite{Edwa} states that given a function $f (x)$ that in a neighborhood of the origin 
admits a series expansion of the form
\begin{equation}
\label{eq:exp}
f (x) = \sum_{n = 0}^\infty \varphi(n)\,\frac{(- 1)^n}{n!}\,x^n\,,
\end{equation}
with $\varphi (0) \neq 0$\footnote{For $\nu > 0$, this condition guarantees the convergence of the integral near $x = 0$.}, 
then
\begin{equation}
\label{eq:rmt}
\int_0^\infty \de x\, x^{\nu - 1}\,f(x) = \Gamma (\nu)\,\varphi (- \nu)\,, \qquad\qquad (\nu \in \mathbb{R})\,.
\end{equation}
In a recent paper \cite{Gonz}, the usefulness of the RMT for the evaluation of integrals associated to some Feynman diagrams 
has been emphasized. 

The use of the formalism of the umbral calculus \cite{Roma} 
\begin{equation}
\label{eq:umbr}
\varphi (n) = \hat{c}^{\,n}\,\varphi (0) 
\end{equation}
allows to cast eq.\eqref{eq:exp} in the form
\begin{equation}
f (x) = \e^{- \hat{c}\,x}\,\varphi (0)\,,
\end{equation}
and, since eq.\eqref{eq:umbr} holds $\forall n \in \mathbb{R}$, we formally get
\begin{equation}
\label{eq:iden}
\int_0^\infty \de x\, x^{\nu - 1}\,\e^{- \hat{c}\,x}\,\varphi (0) = \Gamma (\nu)\,\hat{c}^{- \nu}\,\varphi (0) = 
\Gamma (\nu)\,\varphi (- \nu)\,.
\end{equation}
The procedure we have followed is not a rigorous proof but just an operational ``tool" useful to formulate the RMT. We follow 
therefore the fairly pragmatic criterion of exploiting the umbral method to get a generalization of the RMT integration 
formula and then check it a posteriori for a number of specific examples.

Let us consider the case of a function $f (x)$ that admits the following series expansion around the origin $(m \in \mathbb{N})$
\begin{equation}
f (x) = \sum_{n = 0}^\infty \varphi(n)\,\frac{(- 1)^n}{n!}\,(x^m)^n
\end{equation}
with $f (0) = \varphi (0) \neq 0$. Along the same lines followed for proving the identity \eqref{eq:iden}, it is easy to show that
\begin{eqnarray}
\label{eq:resu}
\int_0^\infty \de x\, x^{\nu - k}\,f(x) \ug \frac1m\,\Gamma \left(\frac{\nu + 1 - k}m\right)\,\hat{c}^{- \frac{\nu + 1 - k}m}\,\varphi (0) \nn \\
\ug \frac1m\,\Gamma \left(\frac{\nu + 1 - k}m\right)\,\varphi \left(-\frac{\nu + 1 - k}m\right)\,,
\end{eqnarray}
where, in this case, the convergence near to $x = 0$ requires $\nu > k - 1$. 

This result is clearly a conjecture, waiting for a more solid justification, but it has been checked numerically for different values of integers 
$m$ and $k$, and for different types of functions $f (x)$.  As an example of application of eq.\eqref{eq:resu}, in the case $\nu = k$, $m = 2$ 
we get 
\begin{equation}
\int_{- \infty}^\infty \de x\,f(x) = \sqrt{\pi}\,\hat{c}^{- 1/2}\,\varphi (0) = \sqrt{\pi}\,\varphi (- 1/2)\,.
\end{equation}

Further discussions will be presented elsewhere in a more detailed investigation.

\end{document}